\documentclass[preprint]{ptephy_v1}

\preprintnumber{KYUSHU-HET-283, OU-HET-1224} 

\usepackage{graphics}

\usepackage{amsmath} 
\usepackage{amsthm} 
\usepackage{url} 
\usepackage{stmaryrd}
\usepackage{tikz}
\usepackage{physics}

\newcommand{\Slash}[1]{{\ooalign{\hfil/\hfil\crcr$#1$}}}
\numberwithin{equation}{section}

\allowdisplaybreaks

\begin{document}

\title{Yet another lattice formulation of 2D $U(1)$ chiral gauge theory via
bosonization}


\author{Okuto Morikawa}
\affil{Department of Physics, Osaka University, Toyonaka, Osaka 560-0043,
Japan}

\author[2]{Soma Onoda}

\author[2]{Hiroshi Suzuki}

\affil[2]{Department of Physics, Kyushu University, 744 Motooka, Nishi-ku,
Fukuoka 819-0395, Japan}





\begin{abstract}%
Recently, lattice formulations of Abelian chiral gauge theory in two
dimensions have been devised on the basis of the Abelian bosonization. A
salient feature of these 2D lattice formulations is that the gauge invariance
is \emph{exactly\/} preserved for anomaly-free theories and thus is completely
free from the question of the gauge mode decoupling. In the present paper, we
propose a yet another lattice formulation sharing this desired property. A
particularly unique point in our formulation is that the vertex operator of the
dual scalar field, which carries the vector charge of the fermion and the
``magnetic charge'' in the bosonization, is represented by a ``hole'' excised
from the lattice; this is the excision method formulated recently by Abe et al.
in a somewhat different context.
\end{abstract}

\subjectindex{B01, B05, B31, B34}

\maketitle

\section{Introduction and conclusion}
\label{sec:1}
Non-perturbative definition of chiral gauge theory is still quite difficult
and, considering its principal and practical importance, it should be studied
from various perspectives~\cite{Luscher:2000hn,Kaplan:2009yg}. Recently,
lattice formulations of Abelian chiral gauge theory in two dimensions (2D) have
been devised~\cite{DeMarco:2023hoh,Berkowitz:2023pnz} on the basis of the
Abelian bosonization~\cite{Coleman:1974bu,Mandelstam:1975hb} (further basic
references on the bosonization are reprinted in~Ref.~\cite{Stone:1995ys}; \S7.5
of~Ref.~\cite{Tong} provides a very precise and concise exposition). A salient
feature of these 2D lattice formulations is that the gauge invariance is
\emph{exactly\/} preserved for anomaly-free theories even before taking the
continuum limit. This is a very important achievement because if a lattice
formulation preserves the exact gauge invariance, then the gauge mode
completely decouples under the lattice regularization and, assuming the
chirality projection on the correct degrees of freedom is also archived, it
becomes quite conceivable that the system is in the same universality class as
a target anomaly-free chiral gauge theory. In four dimensions, this kind of
exact lattice gauge invariance for anomaly-free chiral gauge theories has been
archived only for $U(1)$~\cite{Luscher:1998du} and the electroweak
$SU(2)\times U(1)$~\cite{Kikukawa:2000kd,Kadoh:2007xb} cases. Although the
former lattice formulation can be applied also to the 2D $U(1)$ chiral gauge
theory which we will study here, one finds that the mechanism for the exact
lattice gauge invariance is much simpler here; this is because of the fact that
the quantum anomaly is reproduced in the classical level in bosonization.

A tricky point in the above lattice formulations on the basis of bosonization
is how to represent the vertex operator~$e^{i\Tilde{\phi}}$ of the dual scalar
field~$\Tilde{\phi}$, which is roughly related to the original $2\pi$ periodic
scalar field~$\phi$ by~$\star\dd\phi\sim2\dd\Tilde{\phi}$. This vertex operator
carries a ``magnetic charge'' associated with the conserved ``magnetic
current'' $j^{(m)}:=\dd\phi/2\pi$. In the bosonization, this magnetic object is
necessary to represent an operator which carries the vector charge of the
fermion, such as the fermion field itself. In particular, it is necessary to
construct fermion zero modes which saturate a non-zero index in topologically
non-trivial sectors. Since the magnetic current trivially conserves
$\dd j^{(m)}=\dd^2\phi/2\pi=0$ when $\phi$ is smooth, in order to endow a
non-zero magnetic charge to the vertex operator, one has to assume that $\phi$
is somehow singular at the position of the vertex operator. The most notable
difference between the two formulations
in~Refs.~\cite{DeMarco:2023hoh,Berkowitz:2023pnz} is attributed to, to our
understanding, how to implement this singular nature on the lattice.

In the present paper, we propose yet another lattice formulation sharing the
exact lattice gauge invariance for anomaly-free theories on the basis of
bosonization. The main difference of our formulation
from~Refs.~\cite{DeMarco:2023hoh,Berkowitz:2023pnz} is again how to implement
the above singular nature (the breaking of the Bianchi identity) at the
position of the vertex operator of the dual scalar field. For this, we employ
the ``excision method'' recently formulated by Abe et al.~\cite{Abe:2023uan}
under a somewhat different motivation.\footnote{The original motivation
of~Ref.~\cite{Abe:2023uan} is to formulate the interplay between the
topological charge in the presence of the $\mathbb{Z}_N$ 2-form gauge field on
the lattice~\cite{Abe:2022nfq,Abe:2023ncy} and the 't~Hooft line operator. Such
a formulation would enable one to carry out the analyses in, e.g.\
Ref.~\cite{Gaiotto:2017yup} straightforwardly in a lattice regularized
framework.} In this method, a smoothness of lattice scalar field (the so-called
admissibility~\cite{Luscher:1981zq,Hernandez:1998et,Luscher:1998du}) which
ensures the conservation of the magnetic current (the Bianchi identity) and
the breaking of the Bianchi identity at the magnetic object are reconciled by
excising a ``hole'' on the lattice. In this way, one can define a vertex
operator of the dual scalar field with desired quantized charges. The details
of the excision method will be recapitulated below.

Now, in terms of bosonization, the fermion sector of a 2D $U(1)$ chiral gauge
theory is represented by the continuum action\footnote{We basically follow the
notational convention of~Ref.~\cite{Abe:2023uan}. The compactification
radius~$R$ in~Ref.~\cite{Abe:2023uan} is taken to be~$R=1/\sqrt{2}$ in order to
represent 2D fermions.}
\begin{equation}
   S_{\mathrm{B}}=\sum_\alpha \left[
   \frac{R^2}{4\pi}\int_{M_2}\left|\dd\phi_\alpha+2q_{A,\alpha}A\right|^2
   +\frac{i}{2\pi}q_{V,\alpha}
   \int_{M_2}A\wedge\left(\dd\phi_\alpha+2q_{A,\alpha}A\right)
   \right],
\label{eq:(1.1)}
\end{equation}
where $\alpha=1$, \dots, $N_f$ labels the flavor degrees of freedom and
$\phi_\alpha$ are $2\pi$~periodic real scalar fields; $A$ is the $U(1)$ gauge
potential. In order to be compatible with the $2\pi$ periodicity
of~$\phi_\alpha$, parameters $2q_{A,\alpha}$ and $q_{V,\alpha}$ must be integers.
According to the bosonization rule, this system describes a 2D $U(1)$ chiral
gauge theory in which the covariant derivative of the fermions is given
by\footnote{Strictly speaking, the bosonization~\eqref{eq:(1.1)} is equivalent
to the fermion theory in which all possible spin structures are summed over. If
one wants to single out a particular spin structure, one has to introduce a
$\mathbb{Z}_2$ gauge field and the Arf invariant
term~\cite{Thorngren:2018bhj,Karch:2019lnn}; we do not consider these elements
in this paper.}
\begin{align}
   \Slash{D}&=\Slash{\partial}+i\Slash{A}(q_{V,\alpha}+q_{A,\alpha}\gamma_5)
\notag\\
   &=\Slash{\partial}+i\Slash{A}(q_{R,\alpha}P_R+q_{L,\alpha}P_L),
\label{eq:(1.2)}
\end{align}
where
\begin{equation}
   P_{R,L}:=\frac{1\pm\gamma_5}{2}.
\label{eq:(1.3)}
\end{equation}
In particular, it turns out that the ``electric charge'' in the scalar
theory~\eqref{eq:(1.1)}, $2q_{A,\alpha}$, is twice the axial charge of the
fermion, $q_{A\alpha}$. The $U(1)$ charges introduced above are related to each
other as
\begin{equation}
   q_{V,\alpha}=\frac{1}{2}\left(q_{R,\alpha}+q_{L,\alpha}\right),\qquad
   q_{A,\alpha}=\frac{1}{2}\left(q_{R,\alpha}-q_{L,\alpha}\right).
\label{eq:(1.4)}
\end{equation}
Since the gauge anomaly in the target chiral gauge theory is proportional to
\begin{equation}
   \sum_{\alpha}\left(q_{R,\alpha}^2-q_{L,\alpha}^2\right)
   =4\sum_{\alpha}q_{V,\alpha}q_{A,\alpha},
\label{eq:(1.5)}
\end{equation}
the anomaly cancellation condition among flavors can be written as
\begin{equation}
   \sum_\alpha q_{V,\alpha}q_{A,\alpha}=0.
\label{eq:(1.6)}
\end{equation}

Since we have assumed that $q_{V,\alpha}$ and~$2q_{A,\alpha}$
in~Eq.~\eqref{eq:(1.1)} are integers, our present lattice formulation is
applicable only to cases in which $q_{V,\alpha}$ given by~Eq.~\eqref{eq:(1.4)}
are integers. These include the so-called $345$~model, in which $q_{R,1}=5$,
$q_{L,1}=3$, $q_{R,2}=0$, $q_{L,2}=4$ (i.e., $q_{V,1}=4$, $q_{A,1}=1$, $q_{V,2}=2$,
$q_{A,2}=-2$), but not the $21111$~model, in which $q_{R,1}=0$, $q_{L,1}=2$, and
$q_{R,\alpha}=1$, $q_{L,\alpha}=0$ for~$\alpha=2$--5, because some
of~$q_{V,\alpha}$ become half-integer. This is a limitation of the present
approach on the basis of the bosonization.

Our lattice formulation of 2D chiral gauge theory according to the above idea
is presented in the next section. This is still a theoretical clarification and
it is not obvious if the formulation is amenable to numerical simulations. For
this, it is interesting to investigate a possible equivalent representation
which is free from the sign problem, such as the one presented
in~Ref.~\cite{Berkowitz:2023pnz}. Also, possible generalizations to the
non-Abelian gauge group via the non-Abelian bosonization~\cite{Witten:1983ar}
and to higher dimensional chiral gauge theories are of great interest.

\section{Lattice formulation}
\label{sec:2}
\subsection{Lattice action: scalar part}
\label{sec:2.1}
We consider $2\pi$ periodic real scalar fields on a 2D square lattice~$\Gamma$.
The lattice sites of~$\Gamma$ are denoted as $n$, $m$, etc. For a moment, we
assume that $\Gamma$ is a 2D torus and the scalar fields $e^{i\phi_\alpha(n)}$
obey periodic boundary conditions. We parametrize $e^{i\phi_\alpha(n)}$ by the
logarithm in the principal branch:
\begin{equation}
   \phi_\alpha(n):=\frac{1}{i}\ln e^{i\phi_\alpha(n)},\qquad
   -\pi<\phi_\alpha(n)\leq\pi.
\label{eq:(2.1)}
\end{equation}
This definition of~$\phi_\alpha(n)$ is obviously invariant under the
``$2\pi$~shift,'' $e^{i\phi_\alpha(n)}\to e^{i\phi_\alpha(n)+2\pi i\mathbb{Z}}$.

Our lattice action for the scalar field~\cite{Abe:2023uan} is almost a literal
transcription of the continuum action~\eqref{eq:(1.1)},\footnote{Lorentz
indices $\mu$, $\nu$, etc.\ run over 1 and~2.}
\begin{align}
   S_{\mathrm{B}}&=\sum_\alpha\sum_{n\in\Gamma}
   \Biggl[
   \frac{R^2}{4\pi}\sum_\mu
   D\phi_\alpha(n,\mu)D\phi_\alpha(n,\mu)
   +\frac{i}{2\pi}q_{V,\alpha}\sum_{\mu,\nu}
   \epsilon_{\mu\nu}A_\mu(\Tilde{n})D\phi_\alpha(n+\Hat{\mu},\nu)
\notag\\
   &\qquad\qquad\qquad{}
   +\frac{i}{2}q_{V,\alpha}\sum_{\mu,\nu}
   \epsilon_{\mu\nu}N_{\mu\nu}(\Tilde{n})\phi_\alpha(n+\Hat{\mu}+\Hat{\nu})
   \Biggr],
\label{eq:(2.2)}
\end{align}
except for the last ``counter term,'' which plays a crucial role to lead an
't~Hooft anomaly in the picture of~Ref.~\cite{Abe:2023uan} and the simple form
of the lattice gauge anomaly in the present context. For the continuum limit,
the radius $R^2$ should be tuned to a specific value, not the classical
value~$1/2$ (see, e.g.\ Ref.~\cite{Janke} for this issue).

The explanation of various quantities in this expression is in order:

First, the scalar fields~$\phi_\alpha(n)$ are coupled to a $U(1)$ gauge field;
at this moment, the gauge field is regarded as background and later we will
make it dynamical. For a technical reason caused by introducing the excision
method at a later stage, we first introduce two $U(1)$ lattice gauge fields
(link variables), one on~$\Gamma$ and another on the dual
lattice~$\Tilde{\Gamma}$, whose sites are defined by
\begin{equation}
   \Tilde{n}:=n+\frac{1}{2}\Hat{1}+\frac{1}{2}\Hat{2},
\label{eq:(2.3)}
\end{equation}
as
\begin{equation}
   U(n,\mu)\in U(1),\qquad U(\Tilde{n},\mu)\in U(1).
\label{eq:(2.4)}
\end{equation}
See~Fig.~\ref{fig:1}. Then the latter gauge field is simply given as a copy of
the former:
\begin{equation}
   U(\Tilde{n},\mu):=U(n,\mu).
\label{eq:(2.5)}
\end{equation}
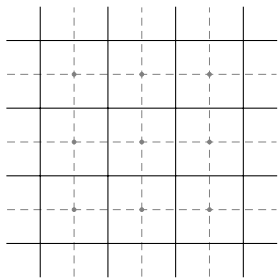
\begin{figure}[htbp]
\centering
\begin{tikzpicture}[scale=0.9]
  \draw[style=help lines,densely dashed,shift={(0.5,0.5)}] (-0.99,-0.99) grid[step=1] (2.99,2.99);
  \foreach \x in {0.5,1.5,2.5} {
    \foreach \y in {0.5,1.5,2.5} {
      \node[draw,circle,style=help lines,inner sep=0.5pt,fill] at (\x,\y) {};
    }
  }
  \draw (-0.5,-0.5) grid[step=1] (3.5,3.5);
\end{tikzpicture}
\caption{The square lattice~$\Gamma$ (the sold line) and the dual
lattice~$\Tilde{\Gamma}$ (the broken lines).}
\label{fig:1}
\end{figure}
Thus, introducing two gauge fields is meaningless at this stage but later when
we introduce a magnetic object by the excised method~\cite{Abe:2023uan}, the
correspondence~\eqref{eq:(2.3)} between the dual and original lattices and,
correspondingly, the relation~\eqref{eq:(2.5)}, must be modified. We
parametrize the link variables by the logarithm in the principal branch:
\begin{align}
   A_\mu(n)&:=\frac{1}{i}\ln U(n,\mu),\qquad
   -\pi<A_\mu(n)\leq\pi,
\notag\\
   A_\mu(\Tilde{n})&:=\frac{1}{i}\ln U(\Tilde{n},\mu),\qquad
   -\pi<A_\mu(\Tilde{n})\leq\pi.
\label{eq:(2.6)}
\end{align}
Of course, $A_\mu(\Tilde{n})=A_\mu(n)$ at this stage.

The directional covariant difference, $D\phi_\alpha(\Tilde{n},\mu)$
in~Eq.~\eqref{eq:(2.2)}, is defined by using the link variable on the
\emph{original\/} lattice, $U(n,\mu)$, as
\begin{align}
   D\phi_\alpha(n,\mu)
   &:=
   \frac{1}{i}\ln\left[
   e^{-i\phi_\alpha(n)}U(n,\mu)^{2q_{A,\alpha}}e^{i\phi_\alpha(n+\Hat{\mu})}\right]
\notag\\
   &=\Delta_\mu\phi_\alpha(n)+2q_{A,\alpha}A_\mu(n)
   +2\pi\ell_{\alpha,\mu}(n),
\label{eq:(2.7)}
\end{align}
where $\Hat{\mu}$ denotes the unit vector in the positive $\mu$-direction and
the branch of the logarithm is taken as~$-\pi<D\phi_\alpha(n,\mu)\leq\pi$. Note
that, when expressed in terms of the simple difference operator,
\begin{equation}
   \Delta_\mu f(n):=f(n+\hat{\mu})-f(n),
\label{eq:(2.8)}
\end{equation}
we generally have the integer field~$\ell_{\alpha,\mu}(n)\in\mathbb{Z}$ as
in~Eq.~\eqref{eq:(2.7)}; $\ell_{\alpha,\mu}(n)$ is a local functional
of~$\phi_\alpha(n)$. By construction, the covariant difference~\eqref{eq:(2.7)}
is \emph{invariant\/} under the lattice gauge transformation:
\begin{align}
   \phi_\alpha(n)
   &\to\phi_\alpha(n)-2q_{A,\alpha}\Lambda(n),
\notag\\
   U(n,\mu)&\to e^{-i\Lambda(n)}U(n,\mu)e^{i\Lambda(n+\Hat{\mu})},
\notag\\
   U(\Tilde{n},\mu)&\to e^{-i\Lambda(\Tilde{n})}U(\Tilde{n},\mu)
   e^{i\Lambda(\Tilde{n}+\Hat{\mu})}.
\label{eq:(2.9)}
\end{align}
Because of~Eq.~\eqref{eq:(2.5)}, the gauge transformation functions are also
identified as~$\Lambda(\Tilde{n})=\Lambda(n)$ at this stage.
Under~Eq.~\eqref{eq:(2.9)}, we see that the gauge potentials
in~Eq.~\eqref{eq:(2.6)} are transformed as
\begin{align}
   A_\mu(n)
   &\to A_\mu(n)+\Delta_\mu\Lambda(n)+2\pi L_\mu(n),
\notag\\
   A_\mu(\Tilde{n})
   &\to A_\mu(\Tilde{n})
   +\Delta_\mu\Lambda(\Tilde{n})+2\pi L_\mu(\Tilde{n}),
\label{eq:(2.10)}
\end{align}
where we have integer fields $L_\mu(n)\in\mathbb{Z}$
and~$L_\mu(\Tilde{n})\in\mathbb{Z}$; $L_\mu(n)$ ($L_\mu(\Tilde{n})$) is a
local functional of~$\Lambda(n)$ ($\Lambda(\Tilde{n})$). Corresponding to the
former, from the gauge invariance of~Eq.~\eqref{eq:(2.7)}, we have the
transformation law of~$\ell_{\alpha,\mu}(n)$:
\begin{equation}
   \ell_{\alpha,\mu}(n)\to\ell_{\alpha,\mu}(n)-2q_{A,\alpha}L_\mu(n).
\label{eq:(2.11)}
\end{equation}

Finally, we introduce gauge invariant field strengths by
\begin{align}
   F_{\mu\nu}(n)
   &:=
   \frac{1}{i}\ln\left[
   U(n,\mu)U(n+\Hat{\mu},\nu)U(n+\Hat{\nu},\mu)^{-1}U(n,\nu)^{-1}\right],
\notag\\
   F_{\mu\nu}(\Tilde{n})
   &:=
   \frac{1}{i}\ln\left[
   U(\Tilde{n},\mu)U(\Tilde{n}+\Hat{\mu},\nu)U(\Tilde{n}+\Hat{\nu},\mu)^{-1}
   U(\Tilde{n},\nu)^{-1}\right],
\label{eq:(2.12)}
\end{align}
where the branch of the logarithm is taken
as~$-\pi<F_{\mu\nu}(n)\leq\pi$ and~$-\pi<F_{\mu\nu}(\Tilde{n})\leq\pi$. Then,
as per~Eq.~\eqref{eq:(2.7)}, we have
\begin{align}
   F_{\mu\nu}(n)
   &=\Delta_\mu A_\nu(n)-\Delta_\nu A_\mu(n)+2\pi N_{\mu\nu}(n),
\notag\\
   F_{\mu\nu}(\Tilde{n})
   &=\Delta_\mu A_\nu(\Tilde{n})-\Delta_\nu A_\mu(\Tilde{n})
   +2\pi N_{\mu\nu}(\Tilde{n}),
\label{eq:(2.13)}
\end{align}
where the integer fields, $N_{\mu\nu}(n)$ and~$N_{\mu\nu}(\Tilde{n})$, are local
functionals of $U(n,\mu)$ and~$U(\Tilde{n},\mu)$, respectively. We note that
under the gauge transformation~\eqref{eq:(2.10)}, $N_{\mu\nu}(n)$
and~$N_{\mu\nu}(\Tilde{n})$ transform as
\begin{align}
   N_{\mu\nu}(n)
   &\to
   N_{\mu\nu}(n)-\Delta_\mu L_\nu(n)+\Delta_\nu L_\mu(n),
\notag\\
   N_{\mu\nu}(\Tilde{n})
   &\to
   N_{\mu\nu}(\Tilde{n})-\Delta_\mu L_\nu(\Tilde{n})+\Delta_\nu L_\mu(\Tilde{n}).
\label{eq:(2.14)}
\end{align}
This completes the explanation on quantities appearing in the lattice
action~\eqref{eq:(2.2)}.

\subsection{Admissibility and the magnetic charge}
\label{sec:2.2}
As the correspondence between~Eqs.~\eqref{eq:(1.1)} and~\eqref{eq:(1.2)} in
continuum theory shows, in the bosonization, the axial-vector current is given
by the ``magnetic'' current~$\epsilon_{\mu\nu}\partial_\nu\phi_\alpha$. Its
lattice counterpart,
\begin{equation}
   \epsilon_{\mu\nu}D\phi_\alpha(n,\nu)
\label{eq:(2.15)}
\end{equation}
however, does not generally conserve,
$\epsilon_{\mu\nu}\Delta_\mu D\phi_\alpha(n,\nu)\neq0$, even when~$A_\mu(n)=0$,
because of the integer field~$\ell_{\alpha,\mu}(n)$ in~Eq.~\eqref{eq:(2.7)},
i.e.\ generally $\epsilon_{\mu\nu}\Delta_\mu\ell_{\alpha,\nu}(n)\neq0$. Then, we
cannot define a conserved axial charge even if the target chiral gauge theory
is anomaly-free. This failure can be evaded by imposing a certain smoothness
condition on lattice fields as follows~\cite{Abe:2023uan}.

We require that possible configurations of the lattice fields satisfy the
following gauge-invariant conditions (the latter two are the
admissibility~\cite{Luscher:1981zq,Luscher:1998du}) for all flavor~$\alpha$:
\begin{equation}
   \sup_{n,\mu}\left|D\phi_\alpha(n,\mu)\right|<\epsilon,\qquad
   \sup_{n,\mu,\nu}\left|2q_{A,\alpha}F_{\mu\nu}(n)\right|<\delta,\qquad
   \sup_{\Tilde{n},\mu,\nu}\left|q_{V,\alpha}F_{\mu\nu}(\Tilde{n})\right|<\delta,
\label{eq:(2.16)}
\end{equation}
where
\begin{equation}
   0<\epsilon<\frac{\pi}{2},\qquad
   0<\delta<\min(\pi,2\pi-4\epsilon).
\label{eq:(2.17)}
\end{equation}
Then, from Eqs.~\eqref{eq:(2.7)} and~\eqref{eq:(2.13)}, these conditions
follow the bound~\cite{Abe:2023uan},
\begin{align}
   &\left|
   \Delta_\mu\ell_{\alpha,\nu}(n)-\Delta_\nu\ell_{\alpha,\mu}(n)
   -2q_{A,\alpha}N_{\mu\nu}(n)
   \right|
\notag\\
   &=\frac{1}{2\pi}
   \left|
   \Delta_\mu D\phi_\alpha(n,\nu)-\Delta_\nu D\phi_\alpha(n,\mu)
   -2q_{A,\alpha}F_{\mu\nu}(n)
   \right|
\notag\\
   &<\frac{2}{\pi}\epsilon+\frac{1}{2\pi}\delta<1.
\label{eq:(2.18)}
\end{align}
Since the leftmost term is a sum of integers, this bound implies that
\begin{align}
   \Delta_\mu\ell_{\alpha,\nu}(n)-\Delta_\nu\ell_{\alpha,\mu}(n)
   &=2q_{A,\alpha}N_{\mu\nu}(n),
\notag\\
   \Delta_\mu D\phi_\alpha(n,\nu)-\Delta_\nu D\phi_\alpha(n,\mu)
   &=2q_{A,\alpha}F_{\mu\nu}(n).
\label{eq:(2.19)}
\end{align}
These are Bianchi identities in the presence of the gauge interaction. Note
that these are invariant under the gauge transformation,
Eqs.~\eqref{eq:(2.11)}, \eqref{eq:(2.14)}, \eqref{eq:(2.9)}
and~\eqref{eq:(2.10)}.

We now define a \emph{gauge-invariant\/} magnetic charge inside a loop~$C$
on~$\Gamma$ by:\footnote{If the gauge field is turned off, this is the magnetic
charge defined in~Ref.~\cite{Abe:2023uan}.}
\begin{align}
   m_\alpha&:=\frac{1}{2\pi}\sum_{(n,\mu)\in C}D\phi_\alpha(n,\mu)
   -\frac{2q_{A,\alpha}}{2\pi}F(C)
\notag\\
   &=\sum_{(n,\mu)\in C}\ell_{\alpha,\mu}(n)
   -2q_{A,\alpha}N(C)\in\mathbb{Z},
\label{eq:(2.20)}
\end{align}
where the sum is taken along the loop. In this expression, we have introduced a
logarithm of the Wilson loop along~$C$,
\begin{equation}
   F(C):=\frac{1}{i}\ln
   \left[
   \prod_{(n,\mu)\in C}U(n,\mu)
   \right]
   =\sum_{(n,\mu)\in C}A_\mu(n)+2\pi N(C),
\label{eq:(2.21)}
\end{equation}
where the branch is $-\pi<F(C)\leq\pi$ and $N(C)$ is an integer.
In deriving the last expression of~Eq.~\eqref{eq:(2.20)}, we have recalled
Eq.~\eqref{eq:(2.7)}; from this, $m_\alpha$ is obviously an integer.

\begin{figure}[htbp]
\centering
\begin{tikzpicture}[scale=0.9]
  \draw[style=help lines,densely dashed,shift={(0.5,0.5)}] (-0.99,-0.99) grid[step=1] (5.99,5.99);
  \foreach \x in {0.5,1.5,2.5,3.5,4.5,5.5} {
    \foreach \y in {0.5,1.5,2.5,3.5,4.5,5.5} {
      \node[draw,circle,style=help lines,inner sep=0.5pt,fill] at (\x,\y) {};
    }
  }
  \draw (-0.5,-0.5) grid[step=1] (6.5,6.5);
  \fill[black!0] (2,1) -- (3,1) -- (3,2) -- (5,2) -- (5,4) -- (4,4) -- (4,5) -- (2,5) -- (2,1);
  \foreach \x / \y in {3.5/2, 5/3.5, 3.5/5, 2/3.5} {
    \draw[style=help lines,densely dashed] (\x,\y) -- (3.5,3.5);
  }
  \foreach \x / \y / \ax / \ay / \bx / \by in {3/1.5/2.7/1.5/2.9/3, 2.5/1/2.5/2.2/2.8/3, 2/1.5/2.2/1.5/2.5/3.1, 4.5/2/4.5/2.3/3.5/3.4, 5/2.5/4.7/2.5/3.6/3.5, 4.5/4/4.5/3.7/3.6/3.5, 4/4.5/3.7/4.5/3.5/3.6, 2.5/5/2.5/4.7/3.5/3.6, 2/4.5/2.3/4.5/3.4/3.5, 2/2.5/2.1/2.5/2.4/3.2,2/1.5/2.2/1.5/2.5/3.1
} {
    \draw[style=help lines,thin,densely dashed] (\x,\y) .. controls (\ax,\ay) and (\bx,\by) .. (3.5,3.5);
  }
  \draw[ultra thick] (2,1) -- (3,1) -- (3,2) -- (5,2) -- (5,4) -- (4,4) -- (4,5) -- (2,5) -- (2,4)
-- (2,1);
  \fill (3.5,3.5) circle(0.1) node[below] {$\Tilde{n}_{\ast}$};
  \draw (2.8,4.2) node {$\mathcal{D}$};
\end{tikzpicture}
\caption{An excised region~$\mathcal{D}$ on~$\Gamma$ and the corresponding dual
lattice~$\Tilde{\Gamma}$ (the broken lines). Inside the excised region, we
place a single site of~$\Tilde{\Gamma}$, $\Tilde{n}_*$. Then the dual lattice
in~$\mathcal{D}$ is defined as in the figure.}
\label{fig:2}
\end{figure}
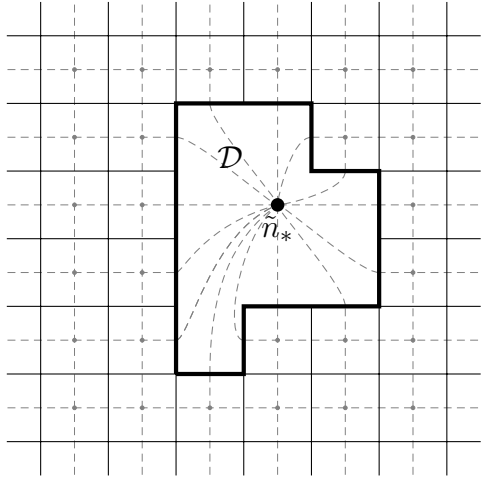

The magnetic charge~$m_\alpha$~\eqref{eq:(2.20)} is conserved in the sense that
it is invariant under any deformation of the loop~$C$ because of the Bianchi
identity~\eqref{eq:(2.19)} (the change due to the integration of the field
strengths is compensated by the change in the Wilson loop). This topological
invariance however implies that $m_\alpha\equiv0$ on a uniform lattice. We may
nevertheless introduce a magnetically-charged object in the system by excising
a certain region~$\mathcal{D}$ from~$\Gamma$, as in~Fig.~\ref{fig:2}. This is
the excision method of~Ref.~\cite{Abe:2023uan}. The magnetic charge of the
excised region~$\mathcal{D}$ is given by setting $C=\partial\mathcal{D}$
in~Eq.~\eqref{eq:(2.20)}. For a later argument, we assume that
for~$C=\partial\mathcal{D}$, for all flavors~$\alpha$,
\begin{equation}
   \left|2q_{A,\alpha}F(\partial\mathcal{D})\right|\leq\delta'
\label{eq:(2.22)}
\end{equation}
with a sufficiently small~$\delta'$ (we will later specify how $\delta'$
should be small). Because of the first condition of~Eq.~\eqref{eq:(2.16)} and
this, the size of the region~$\mathcal{D}$ is of order~$1$ in lattice units and
the magnetic object becomes point-like in the continuum limit; the precise
shape of~$\mathcal{D}$ is thus not important. Note that from the bosonization
rule, the magnetic object carries the vector $U(1)$ charge but no axial $U(1)$
charge. This localized magnetic object can be identified with the vertex
operator~$e^{im_\alpha\Tilde{\phi}_\alpha(x)}$ of the dual scalar
field~$\Tilde{\phi}_\alpha(x)$ in the continuum theory. To realize this
magnetic object in quantum theory, the functional integral
over~$\phi_\alpha(n)$ must be carried out under the boundary
condition~\eqref{eq:(2.20)}.

With the presence of an excised region~$\mathcal{D}$ as depicted
in~Fig.~\ref{fig:2}, however, the correspondence between links of~$\Gamma$ and
those of~$\Tilde{\Gamma}$ is not obvious and the simple identification of gauge
fields~\eqref{eq:(2.5)} does not work. For such a case with an excised region,
we adopt the following rule: If the link~$(\Tilde{n},\mu)$ on~$\Tilde{\Gamma}$
possesses the corresponding link~$(n,\mu)$ on~$\Gamma$
with~Eq.~\eqref{eq:(2.3)}, we apply the simple identification~\eqref{eq:(2.5)}.
We find that precisely when the link~$(\Tilde{n},\mu)$ crosses the boundary
of~$\mathcal{D}$, $\partial\mathcal{D}$ (i.e., when $\Tilde{n}=\Tilde{n}_*$
or~$\Tilde{n}+\Hat{\mu}=\Tilde{n}_*$), such a correspondence breaks. For such a
case, $U(\Tilde{n},\mu)$ is not given by~$U(n,\mu)$ and it should be treated
as an independent variable:
\begin{equation}
   \text{$U(\Tilde{n},\mu)$ is an independent variable if the link
   $(\Tilde{n},\mu)$ crosses the boundary of~$\mathcal{D}$}.
\label{eq:(2.23)}
\end{equation}
Also, the gauge transformation functions, $\Lambda(\Tilde{n})$, can almost be
given by~$\Lambda(n)$ on~$\Gamma$ by~Eq.~\eqref{eq:(2.3)} but we find that
the site~$\Tilde{n}_*$ is the exception, i.e.,
\begin{equation}
   \text{$\Lambda(\Tilde{n}_*)$ is an independent gauge transformation
   variable}.
\label{eq:(2.24)}
\end{equation}

\subsection{Anomaly under the gauge transformation}
\label{sec:2.3}
Now, let us study how our lattice action~\eqref{eq:(2.2)} changes under the
lattice gauge transformation. In general, it should not be invariant in order
to reproduce the gauge anomaly in the target chiral gauge theory. One may still
hope that the lattice action becomes perfectly gauge invariant if the anomaly
cancellation condition~\eqref{eq:(1.6)} matches. Remarkably, this actually
occurs.

It is straightforward to see that under the gauge transformation the
action~$S_{\mathrm{B}}$~\eqref{eq:(2.2)} changes into~\cite{Abe:2023uan}
\begin{align}
   &S_{\mathrm{B}}
   +\frac{i}{2\pi}\sum_\alpha q_{V,\alpha}
   \sum_{n\in\Gamma}\sum_{\mu,\nu}\epsilon_{\mu\nu}
   \Delta_\mu\Lambda(\Tilde{n})D\phi_\alpha(n+\Hat{\mu},\Hat{\nu})
\notag\\
   &\qquad{}
   -i\sum_\alpha q_{V,\alpha}q_{A,\alpha}
   \sum_{n\in\Gamma}\sum_{\mu,\nu}\epsilon_{\mu\nu}
   \left[N_{\mu\nu}(\Tilde{n})
   -\Delta_\mu L_\nu(\Tilde{n})+\Delta_\nu L_\mu(\Tilde{n})\right]
   \Lambda(n+\Hat{\mu}+\Hat{\nu})
\notag\\
   &\qquad{}
   +2i\sum_\alpha q_{V,\alpha}q_{A,\alpha}
   \sum_{n\in\Gamma}\sum_{\mu,\nu}\epsilon_{\mu\nu}
   L_\mu(\Tilde{n})A_\nu(n+\Hat{\mu})
\notag\\
   &\qquad{}
   +i\sum_\alpha q_{V,\alpha}
   \sum_{n\in\Gamma}\sum_{\mu,\nu}\epsilon_{\mu\nu}
   \Delta_\nu\left[
   L_\mu(\Tilde{n})\phi_\alpha(n+\Hat{\mu})\right]
\notag\\
   &\qquad{}
   +2\pi i\sum_\alpha q_{V,\alpha}
   \sum_{n\in\Gamma}\sum_{\mu,\nu}\epsilon_{\mu\nu}
   L_\mu(\Tilde{n})\ell_\nu(n+\Hat{\mu}).
\label{eq:(2.25)}
\end{align}
The last term is~$2\pi i\mathbb{Z}$ and is negligible in the partition
function,
\begin{equation}
   \int\prod_\alpha\prod_{n\in\Gamma}d\phi_\alpha(n)\,e^{-S_{\mathrm{B}}}.
\label{eq:(2.26)}
\end{equation}
Also, the total divergence term~$\Delta_\nu[\dotsc]$ can be neglected because
we are assuming the boundary conditions are periodic.

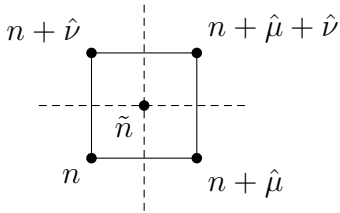
\begin{figure}[htbp]
\centering
\begin{tikzpicture}[scale=0.7]
  \draw[thin] (-1,-1) rectangle (1,1);
  \draw[thin,densely dashed] (-2,0) -- (2,0);
  \draw[thin,densely dashed] (0,-2) -- (0,2);
  \fill (-1,-1) circle(0.1) node[below left]  {$n$};
  \fill (1,-1)  circle(0.1) node[below right] {$n+\Hat{\mu}$};
  \fill (-1,1)  circle(0.1) node[above left]  {$n+\Hat{\nu}$};
  \fill (1,1)   circle(0.1) node[above right] {$n+\Hat{\mu}+\Hat{\nu}$};
  \fill (0,0) circle(0.1) node[below left] {$\Tilde{n}$};
\end{tikzpicture}
\caption{Structure appearing in Eq.~\eqref{eq:(2.27)}.}
\label{fig:3}
\end{figure}

The second term of~Eq.~\eqref{eq:(2.25)},
$\sum_{n\in\Gamma}\sum_{\mu,\nu}\epsilon_{\mu\nu}\Delta_\mu\Lambda(\Tilde{n})D\phi_\alpha(n+\Hat{\mu},\Hat{\nu})$,
requires a close look~\cite{Abe:2023uan}. By considering combinations
containing~$\Lambda(\Tilde{n})$ with a particular site~$\Tilde{n}$
(see~Fig.~\ref{fig:3}), one finds that the second term can be written as
\begin{align}
   &\frac{i}{2\pi}\sum_\alpha q_{V,\alpha}
   \sum_{n\in\Gamma}\sum_{\mu,\nu}\epsilon_{\mu\nu}
   \Delta_\mu\Lambda(\Tilde{n})D\phi_\alpha(n+\Hat{\mu},\Hat{\nu})
\notag\\
   &=-\frac{i}{4\pi}\sum_\alpha q_{V,\alpha}
   \sum_{n\in\Gamma}\sum_{\mu,\nu}\epsilon_{\mu\nu}
   \Lambda(\Tilde{n})
   \left[
   \Delta_\mu D\phi_\alpha(n,\Hat{\nu})-\Delta_\nu D\phi_\alpha(n,\Hat{\mu})
   \right]
\notag\\
   &=-\frac{i}{2\pi}\sum_\alpha q_{V,\alpha}q_{A,\alpha}
   \sum_{n\in\Gamma}\sum_{\mu,\nu}\epsilon_{\mu\nu}
   \Lambda(\Tilde{n})F_{\mu\nu}(n),
\label{eq:(2.27)}
\end{align}
where, in the last step, we have noted~Eq.~\eqref{eq:(2.19)}. In total, the
action changes into
\begin{align}
   &S_{\mathrm{B}}+i\left(\sum_\alpha q_{V,\alpha}q_{A,\alpha}\right)
   \sum_{n\in\Gamma}\sum_{\mu,\nu}\epsilon_{\mu\nu}
\notag\\
   &\qquad{}
   \times\biggl\{
   -\frac{1}{2\pi}\Lambda(\Tilde{n})F_{\mu\nu}(n)
   -
   \left[N_{\mu\nu}(\Tilde{n})
   -\Delta_\mu L_\nu(\Tilde{n})+\Delta_\nu L_\mu(\Tilde{n})\right]
   \Lambda(n+\Hat{\mu}+\Hat{\nu})
\notag\\
   &\qquad\qquad{}+2L_\mu(\Tilde{n})A_\nu(n+\Hat{\mu})
   \biggr\},
\label{eq:(2.28)}
\end{align}
up to~$2\pi i\mathbb{Z}$. Since the change is proportional to the anomaly
coefficient~\eqref{eq:(1.5)}, for anomaly-free chiral gauge theories, our
lattice action and hence the regularized partition function~\eqref{eq:(2.26)}
is \emph{exactly\/} gauge invariant.

The above argument proceeds in an almost equal way, even with an excised
region~$\mathcal{D}$ which represents a magnetically charged object. For this
case, we have the contribution from~$\mathcal{D}$ in addition
to~Eq.~\eqref{eq:(2.28)}~\cite{Abe:2023uan},
\begin{align}
   &-i\sum_\alpha q_{V,\alpha}
   \Lambda(\Tilde{n}_*)
   \sum_{(n,\mu)\in\partial\mathcal{D}}
   \left[
   \ell_{\alpha,\mu}(n)
   +\frac{2q_{A,\alpha}}{2\pi}A_\mu(n)
   \right]
\notag\\
   &=-i\sum_\alpha q_{V,\alpha}
   \Lambda(\Tilde{n}_*)
   \left[
   \sum_{(n,\mu)\in\partial\mathcal{D}}
   \ell_{\alpha,\mu}(n)-2q_{A,\alpha}N(\partial\mathcal{D})
   \right]
   -\frac{i}{\pi}\left(\sum_\alpha q_{V,\alpha}q_{A,\alpha}\right)
   \Lambda(\Tilde{n}_*)
   F(\partial\mathcal{D})
\notag\\
   &=-i\sum_\alpha q_{V,\alpha}m_\alpha\Lambda(\Tilde{n}_*),
\label{eq:(2.29)}
\end{align}
where, in the first equality, we have used Eq.~\eqref{eq:(2.21)}
with~$C=\partial\mathcal{D}$. In the second equality, we have used the
anomaly cancellation condition~\eqref{eq:(1.6)} and the definition of the
magnetic charge~\eqref{eq:(2.20)} taking~$C=\partial\mathcal{D}$. This
breaking of the gauge symmetry owing to the magnetic object may be cured by
connecting to the magnetic object an ``open 't~Hooft line'' in the dual
lattice~\cite{Abe:2023uan},
\begin{equation}
   \exp\left[
   -i\sum_\alpha q_{V,\alpha}m_\alpha
   \sum_{(\Tilde{n},\mu)\in\Tilde{P}}^{\Tilde{n}_*}A_\mu(\Tilde{n})
   \right],
\label{eq:(2.30)}
\end{equation}
where $\Tilde{P}$ denotes a path on the dual lattice ending at~$\Tilde{n}_*$.
In this way, under the gauge anomaly cancellation condition~\eqref{eq:(1.6)},
the fermion sector of the target chiral gauge theory is formulated in a
manifestly gauge invariant manner with the lattice regularization.

\subsection{Selection rule and the fermion zero modes}
\label{sec:2.4}
Under the gauge transformation~\eqref{eq:(2.9)}, the vertex operator of the
original scalar fields,
\begin{equation}
   V_{\{n_\alpha \}}(n):=e^{i\sum_\alpha n_\alpha \phi_\alpha(n)},
\label{eq:(2.31)}
\end{equation}
changes as
\begin{equation}
   V_{\{n_\alpha \}}(n)
   \to\exp\left[-i\sum_\alpha 2q_{A,\alpha}n_\alpha \Lambda(n)\right]
   V_{\{n_\alpha \}}(n).
\label{eq:(2.32)}
\end{equation}
Thus, this vertex operator possesses the axial $U(1)$ charge
$\sum_\alpha2q_{A,\alpha}n_\alpha$ while no vector~$U(1)$ charge. We can make this
vertex operator gauge invariant by connecting an open Wilson line
\begin{equation}
   \exp\left[
   i\sum_\alpha2q_{A,\alpha}n_\alpha
   \sum_{(n,\mu)\in P}^nA_\mu(n)
   \right]
\label{eq:(2.33)}
\end{equation}
to it, where $P$ denotes a path on the lattice ending at the position of the
vertex operator, $n$.

If we perform a constant shift of~$\phi_\alpha(n)$,
$\phi_\alpha(n)\to\phi_\alpha(n)+\xi_\alpha$, in the partition
function~\eqref{eq:(2.26)}, the lattice action~$S_{\mathrm{B}}$~\eqref{eq:(2.2)}
produces a factor,
\begin{equation}
   \exp\left[-i\sum_\alpha \xi_\alpha q_{V,\alpha}
   \sum_{\Tilde{p}\in\Tilde{\Gamma}}N_{12}(\Tilde{p})\right]
   =\exp\left[-\frac{i}{2\pi}\sum_\alpha \xi_\alpha q_{V,\alpha}
   \sum_{\Tilde{p}\in\Tilde{\Gamma}}F_{12}(\Tilde{p})\right],
\label{eq:(2.34)}
\end{equation}
where the lattice sum is taken over all plaquettes on the dual
lattice~$\Tilde{\Gamma}$. This shows that, when the lattice first Chern number
of the background gauge field,
\begin{equation}
   \Tilde{Q}
   :=\frac{1}{2\pi}\sum_{\Tilde{p}\in\Tilde{\Gamma}}F_{12}(\Tilde{p})
   =\sum_{\Tilde{p}\in\Tilde{\Gamma}}N_{12}(\Tilde{p})\in\mathbb{Z}
\label{eq:(2.35)}
\end{equation}
is non-zero, the partition function vanishes as expected from the index theorem
in the corresponding gauge theory containing fermions. Here, we should note
that the lattice first Chern number~\eqref{eq:(2.35)} can be non-trivial under
the admissibility~\eqref{eq:(2.16)}~\cite{Luscher:1998du}. We can make the
partition function~\eqref{eq:(2.26)} non-zero by inserting vertex
operators~\eqref{eq:(2.31)} in an appropriate way. Since
\begin{equation}
   V_{\{n_\alpha \}}(n)
   \to e^{i\sum_\alpha \xi_\alpha n_\alpha}V_{\{n_\alpha \}}(n),
\label{eq:(2.36)}
\end{equation}
under the constant shift, the condition for non-vanishingness is
\begin{equation}
   \sum_In_{I,\alpha}
   =\frac{q_{V,\alpha}}{2\pi}\sum_{\Tilde{p}\in\Tilde{\Gamma}}F_{12}(\Tilde{p})
   =q_{V,\alpha}\Tilde{Q},
\label{eq:(2.37)}
\end{equation}
where $I$ labels the inserted vertex operators.

On the other hand, for the magnetic object introduced by the excision method,
\begin{equation}
   M_{\{m_\alpha \}}(\mathcal{D}),
\label{eq:(2.38)}
\end{equation}
the gauge transformation induces (see~Eq.~\eqref{eq:(2.29)}),
\begin{equation}
   M_{\{m_\alpha \}}(\mathcal{D})
   \to
   \exp\left[
   i\sum_\alpha q_{V,\alpha}m_\alpha \Lambda(\Tilde{n}_*)
   \right]
   M_{\{m_\alpha \}}(\mathcal{D}).
\label{eq:(2.39)}
\end{equation}
Therefore, this object possesses the vector~$U(1)$ charge
$-\sum_\alpha q_{V,\alpha}m_\alpha$ while possessing no axial $U(1)$ charge. For the
magnetic objects, we have the consistency condition,
\begin{equation}
   \sum_{\Tilde{I}}m_{\Tilde{I},\alpha}
   =-\frac{2q_{A,\alpha}}{2\pi}\sum_{p\in\Gamma-\sum_{\Tilde{I}}\mathcal{D}_{\Tilde{I}}}
   F_{12}(p)
   -\frac{2q_{A,\alpha}}{2\pi}\sum_{\Tilde{I}}F(\partial\mathcal{D}_{\Tilde{I}})
   =-2q_{A,\alpha}Q,
\label{eq:(2.40)}
\end{equation}
where $\Tilde{I}$ labels the magnetic objects and the first lattice sum on the
right-hand side is taken over the plaquette~$p$
on~$\Gamma-\sum_{\Tilde{I}}\mathcal{D}_{\Tilde{I}}$. This consistency is obtained
by simply summing the Bianchi identity~\eqref{eq:(2.19)} over the
region~$\Gamma-\sum_{\Tilde{I}}\mathcal{D}_{\Tilde{I}}$ and noting the definition
of the magnetic charge~\eqref{eq:(2.20)}. In Eq.~\eqref{eq:(2.40)}, we have
introduced another lattice first Chern number:
\begin{align}
   Q&:=\frac{1}{2\pi}\sum_{p\in\Gamma-\sum_{\Tilde{I}}\mathcal{D}_{\Tilde{I}}}
   F_{12}(p)
   +\frac{1}{2\pi}\sum_{\Tilde{I}}F(\partial\mathcal{D}_{\Tilde{I}})
\notag\\
   &=\sum_{p\in\Gamma-\sum_{\Tilde{I}}\mathcal{D}_{\Tilde{I}}}N_{12}(p)
   +\sum_{\Tilde{I}}N(\partial\mathcal{D}_{\Tilde{I}})
   \in\mathbb{Z},
\label{eq:(2.41)}
\end{align}
where, in the last line, we have used Eqs.~\eqref{eq:(2.13)}
and~\eqref{eq:(2.21)}. Although we have two lattice Chern numbers
$\Tilde{Q}$~\eqref{eq:(2.35)} and~$Q$~\eqref{eq:(2.41)}, one might expect that
those two coincide in the continuum limit. In fact, since the difference
between $\Tilde{Q}$ and~$Q$ arises only from the differences in the sum
of~$F_{12}(\Tilde{n})$ and of~$F_{12}(n)$ on a \emph{finite\/} number of
plaquettes around the magnetic objects and the sum
of~$F(\partial\mathcal{D}_{\Tilde{I}})$, if we take $\delta$
in~Eq.~\eqref{eq:(2.16)} and $\delta'$ in~Eq.~\eqref{eq:(2.22)} small enough,
the equality of two integers, $\Tilde{Q}=Q$, automatically holds. We assume
this in what follows.

The elementary fermion fields in the target theory may be represented by
``superposing'' the vertex operator~\eqref{eq:(2.31)} and the magnetic
object~\eqref{eq:(2.38)}. For a flavor~$\alpha$, we have
\begin{align}
   &P_R\psi_\alpha:e^{+i\phi_\alpha(n)/2}M_{m_\alpha=-1}(\mathcal{D}),
\notag\\
   &\Bar{\psi}_\alpha P_L:e^{-i\phi_\alpha(n)/2}M_{m_\alpha=+1}(\mathcal{D}).
\notag\\
   &P_L\psi_\alpha:e^{-i\phi_\alpha(n)/2}M_{m_\alpha=-1}(\mathcal{D}),
\notag\\
   &\Bar{\psi}_\alpha P_R:e^{+i\phi_\alpha(n)/2}M_{m_\alpha=+1}(\mathcal{D}).
\label{eq:(2.42)}
\end{align}
These representations can be identified by examining the transformation law
under the gauge transformation. These representations may be used to compute
correlation functions containing fermion fields. These elementary fields are
not gauge invariant but we may dress them by open Wilson and 't~Hooft lines to
render them gauge invariant. For instance,
\begin{equation}
   \widehat{P_L\psi_\alpha}:e^{-i\phi_\alpha(n)/2}M_{m_\alpha=-1}(\mathcal{D})
   \exp\left[
   -iq_{A,\alpha}\sum_{(n,\mu)\in P}^nA_\mu(n)
   +iq_{V,\alpha}
   \sum_{(\Tilde{n},\mu)\in\Tilde{P}}^{\Tilde{n}_*}A_\mu(\Tilde{n})
   \right]
\label{eq:(2.43)}
\end{equation}
is gauge invariant, where the hat~$\widehat{\phantom{x}}$ indicates the field
is dressed. We may use such a gauge invariant field to saturate the selection
rule. When $\Tilde{Q}=Q$, the selection rules in~Eqs.~\eqref{eq:(2.37)}
and~\eqref{eq:(2.40)} imply that correlation functions vanish unless the
(gauge-invariant) $|q_{L,\alpha}Q|$ product of
\begin{equation}
   \begin{cases}
   P_L\psi_\alpha&\text{when $q_{L,\alpha}Q<0$},\\
   \Bar{\psi}_\alpha P_R&\text{when $q_{L,\alpha}Q>0$},\\
   \end{cases}
\label{eq:(2.44)}
\end{equation}
and the (gauge-invariant) $|q_{R,\alpha}Q|$ product of
\begin{equation}
   \begin{cases}
   P_R\psi_\alpha&\text{when $q_{R,\alpha}Q>0$},\\
   \Bar{\psi}_\alpha P_L&\text{when $q_{R,\alpha}Q<0$},\\
   \end{cases}
\label{eq:(2.45)}
\end{equation}
are inserted in addition to pairs of $\psi_\alpha$ and~$\Bar{\psi}_\alpha$. In
the target chiral gauge theory in continuum, one may define gauge-invariant
fermion number currents of the left-handed and right-handed Weyl
fermions, respectively, as (see Ref.~\cite{Fujikawa:1994np} and references
cited therein)
\begin{equation}
   J_\mu^{L,R}(x)
   :=-\lim_{\Lambda\to\infty}\tr\gamma_\mu P_{L,R}\frac{1}{\Slash{D}_{L,R}}
   e^{\Slash{D}_{L,R}^2/\Lambda^2}\delta(x-y)|_{y\to x},
\label{eq:(2.46)}
\end{equation}
where $\Slash{D}_{L,R}:=\gamma_\mu(\partial_\mu+iq_{L,R}A_\mu)$. Then the fermion
number anomaly is given by\footnote{Our convention is
$\gamma_5=i\gamma_1\gamma_2$ and~$\epsilon_{12}=1$.}
\begin{align}
   \partial_\mu J_\mu^{L,R}(x)
   &=\mp\lim_{\Lambda\to\infty}\tr\gamma_5
   e^{\Slash{D}_{L,R}^2/\Lambda^2}\delta(x-y)|_{y\to x}
   =\mp\frac{q_{L,R}}{2\pi}F_{12}(x).
\label{eq:(2.47)}
\end{align}
The above selection rule in~Eqs.~\eqref{eq:(2.44)} and~\eqref{eq:(2.45)} is
precisely consistent with the 2D integral of these non-conservation laws.

\subsection{Making the gauge field dynamical}
\label{sec:2.5}
It is straightforward to make the hitherto background $U(1)$ gauge fields
dynamical while preserving the manifest gauge invariance. The total
expectation value of a gauge-invariant observable~$\mathcal{O}$ may be defined
by the functional integral,
\begin{equation}
   \langle\mathcal{O}\rangle
   =\frac{1}{\mathcal{Z}}
   \int\prod_{\text{link $(n,\mu)\in\Gamma-\sum_{\Tilde{I}}\mathcal{D}_{\Tilde{I}}$}}
   dU(n,\mu)\,
   \int\prod_{\text{link $(\Tilde{n},\mu)$ crossing
   $\sum_{\Tilde{I}}\partial\mathcal{D}_{\Tilde{I}}$}}dU(\Tilde{n},\mu)\,
   e^{-S_\mathrm{G}}
   \left\langle\mathcal{O}\right\rangle_{\mathrm{B}},
\label{eq:(2.48)}
\end{equation}
where $\Tilde{I}$ labels the excised regions (magnetic objects) and
\begin{equation}
   \langle\mathcal{O}\rangle_{\mathrm{B}}
   =w[Q]
   \int\prod_\alpha\prod_{n\in\Gamma-\sum_{\Tilde{I}}\mathcal{D}_{\Tilde{I}}}
   d\phi_\alpha(n)\,
   e^{-S_\mathrm{B}}\,\mathcal{O},
\label{eq:(2.49)}
\end{equation}
is the integration over the scalar fields. The total partition function
$\mathcal{Z}$ is determined by requiring $\langle1\rangle=1$. The integrals
over the $U(1)$ gauge fields are defined by the Haar measure at each link;
recall the corresponding rule in~Eqs.~\eqref{eq:(2.5)}
and~\eqref{eq:(2.24)}. We may adopt a somewhat unconventional gauge
action~$S_{\mathrm{G}}$, imitating that of~Ref.~\cite{Luscher:1998du},
\begin{align}
   S_{\mathrm{G}}&=\frac{1}{2g_0^2}
   \sum_{\text{plaquette $(n,1,2)\in\Gamma-\sum_{\Tilde{I}}\mathcal{D}_{\Tilde{I}}$}}
   \mathcal{L}_{12}(n)
\notag\\
   &\qquad{}
   +\frac{1}{2g_0^2}
   \sum_{\text{plaquette $(\Tilde{n},1,2)\in\Tilde{\Gamma}$ containing a
   link crossing $\sum_{\Tilde{I}}\partial\mathcal{D}_{\Tilde{I}}$}}
   \mathcal{L}_{12}(\Tilde{n})
\label{eq:(2.50)}
\end{align}
with $g_0$ being the bare coupling and
\begin{align}
   \mathcal{L}_{\mu\nu}(n)
   &:=\begin{cases}
   [F_{\mu\nu}(n)]^2\left[1-[q_*F_{\mu\nu}(n)]^2
   /\delta^2\right\}^{-1}&
   \text{if $|q_*F_{\mu\nu}(n)|<\delta$},\\
   \infty&\text{otherwise},\\
   \end{cases}
\notag\\
   \mathcal{L}_{\mu\nu}(\Tilde{n})
   &:=\begin{cases}
   [F_{\mu\nu}(\Tilde{n})]^2\left[1-[\Tilde{q}_*F_{\mu\nu}(\Tilde{n})]^2
   /\delta^2\right\}^{-1}&
   \text{if $|\Tilde{q}_*F_{\mu\nu}(\Tilde{n})|<\delta$},\\
   \infty&\text{otherwise},\\
   \end{cases}
\label{eq:(2.51)}
\end{align}
where $q_*:=\max_\alpha|2q_{A,\alpha}|$
and~$\Tilde{q}_*:=\max_\alpha|q_{V,\alpha}|$, to dynamically impose the
admissibility~\eqref{eq:(2.16)}.\footnote{For compatibility with the
admissibility~\eqref{eq:(2.16)}, it might be better to adopt a somewhat
different form of the kinetic term also for the scalar field:
$[D\phi_\alpha(n,\mu)]^2\{1-[D\phi_\alpha(n,\mu)]^2/\epsilon^2\}^{-1}$
if $|D\phi_\alpha(n,\mu)|<\epsilon$ and~$\infty$ otherwise. Any conclusions so
far are not modified even with this choice. It might also be natural to add
a term such as,
$F(\partial\mathcal{D})^2\{1-[q_*F(\partial\mathcal{D})]^2/\delta^{\prime2}\}^{-1}$ if $|q_*F(\partial\mathcal{D})|<\delta^\prime$ and $\infty$ otherwise, to
implement~Eq.~\eqref{eq:(2.22)}.}

As already noted, under the admissibility~\eqref{eq:(2.16)}, the topological
charge~$\Tilde{Q}$~\eqref{eq:(2.35)} is well-defined and can be non-trivial.
The space of such admissible lattice gauge fields is thus a disjoint union of
topological sectors labeled by~$\Tilde{Q}$~\cite{Luscher:1998du}. There is a
freedom in the relative weight and phase for each topological sector and the
factor~$w[\Tilde{Q}=Q]$ parametrizes this freedom.

\section*{Acknowledgments}
We would like to thank Jun Nishimura, Tetsuya Onogi, and Yuya Tanizaki for
suggestive discussions and useful information.
This work was partially supported by Japan Society for the Promotion of Science
(JSPS) Grant-in-Aid for Scientific Research Grant Numbers JP22KJ2096~(O.M.)
and~JP23K03418~(H.S.).




%



\let\doi\relax









\end{document}